# Speckle-based feedback control of optical dipole trap axial waist position

D.A. Pershin[1], D.A. Kumpilov[1,2], A.E. Rudnev[1,2], G.E. Subbotin[1,3], A.M. Ibrahimov[1,2], I.S. Cojocaru[1,4], V.A. Khlebnikov[1], I.A. Pyrkh[1,2], P.A. Aksentsev[1,4], S.A. Kuzmin[1,2], A.D. Raskatov[1,2], A.K. Zykova[1], V.V. Tsyganok[1], A.V. Akimov[1,2,4]

*[1]Russian Quantum Center, Bolshoy Boulevard 30, building 1, Skolkovo, 121205, Russia*

*[2]Moscow Institute of Physics and Technology, Institutskii pereulok 9, Dolgoprudny, Moscow Region 141701, Russia*

*[3]National Research Nuclear University MEPhI, 31, Kashirskoe Highway, Moscow, 115409, Russia*

*[4]PN Lebedev Institute RAS, Leninsky Prospekt 53, Moscow, 119991, Russia*

*Email: a.akimov@rqc.ru*

Cold neutral atoms is a powerful tool for many experiments ranging from frequency standards and sensing to quantum simulations. Sensitive experiments often demand transferring cold atoms from one vacuum chamber to the other with better optical access and vacuum. In case the transfer is done with a beam waist controlled by a focus-tunable lens, repeatability of the transfer as well as stability of the final position can be below the experimental demands. Here we implement the scheme for stabilization of the axial waist position of the optical dipole trap in the whole range of transfer (38 cm) using speckle patterns, which provides enough stability to achieve Bose-Einstein condensation of around $4\cdot10^4$ thulium atoms in the trap formed by the transport beam and the other beam in the target vacuum volume.

Cold atoms are becoming a powerful tool for many fundamental experiments [1–9] and applications [10–18]. In many practical cases it is important to be able to move atoms from one location to another, often to a chamber with better vacuum and optical access. Such a transport could be done in several ways, namely using magnetic traps [19,20], shifted optical lattices [21,22] or moving optical dipole traps (ODT) [23–25]. Moving ODT is applicable to all the atomic species unlike the magnetic transport and does not require precise adjustment of counter-propagating beam frequency as well as position unlike the shifted optical lattices [21,22]

To move the ODT various techniques could be used, for example translation stages [26], lenses with changeable focus such as Moiré lenses [22] and electrically tunable liquid-based lenses [24]. Flexible focus lens lenses are somewhat easier to use especially in case of large displacements, but the downside of this approach is a risk of possible drifts of controlling the beam elements, which may lead to shift of atomic cloud while it needs to be stationary.

Here we demonstrate method of stabilization of ODT longitudinal waist position via feedback using traditionally believed artefacts of laser beams – speckles. The stabilization is experimentally realized on the focus-tunable lens and a complementary metal-oxide-semiconductor (CMOS) camera. The stabilized beam was successfully exploited to transport and cool to quantum degeneracy an ensemble of thulium atoms with consequent achievement of around $4 \cdot 10^4$ atoms in the condensate.

The stabilization of the beam waist was realized during transport of thulium atoms in the single-beam ODT over a distance of 38 cm between the MOT vacuum chamber and the glass cell, the setup repeating the one presented previously [27]. The atoms in the ODT were loaded in the main chamber (MOT on the Figure 1a) from the magneto-optical trap described elsewhere [27–33]. The transport ODT was realized with a linearly polarized laser beam waist of 37 μm and a wavelength of 1064 nm using a focus-tunable lens Optotune EL-16-40-TC-NIR-20D-1-C and a bi-convex lens with focal length 250 mm. The transport was performed exploiting the developed previously technique with supporting constant magnetic field to avoid additional losses [27].

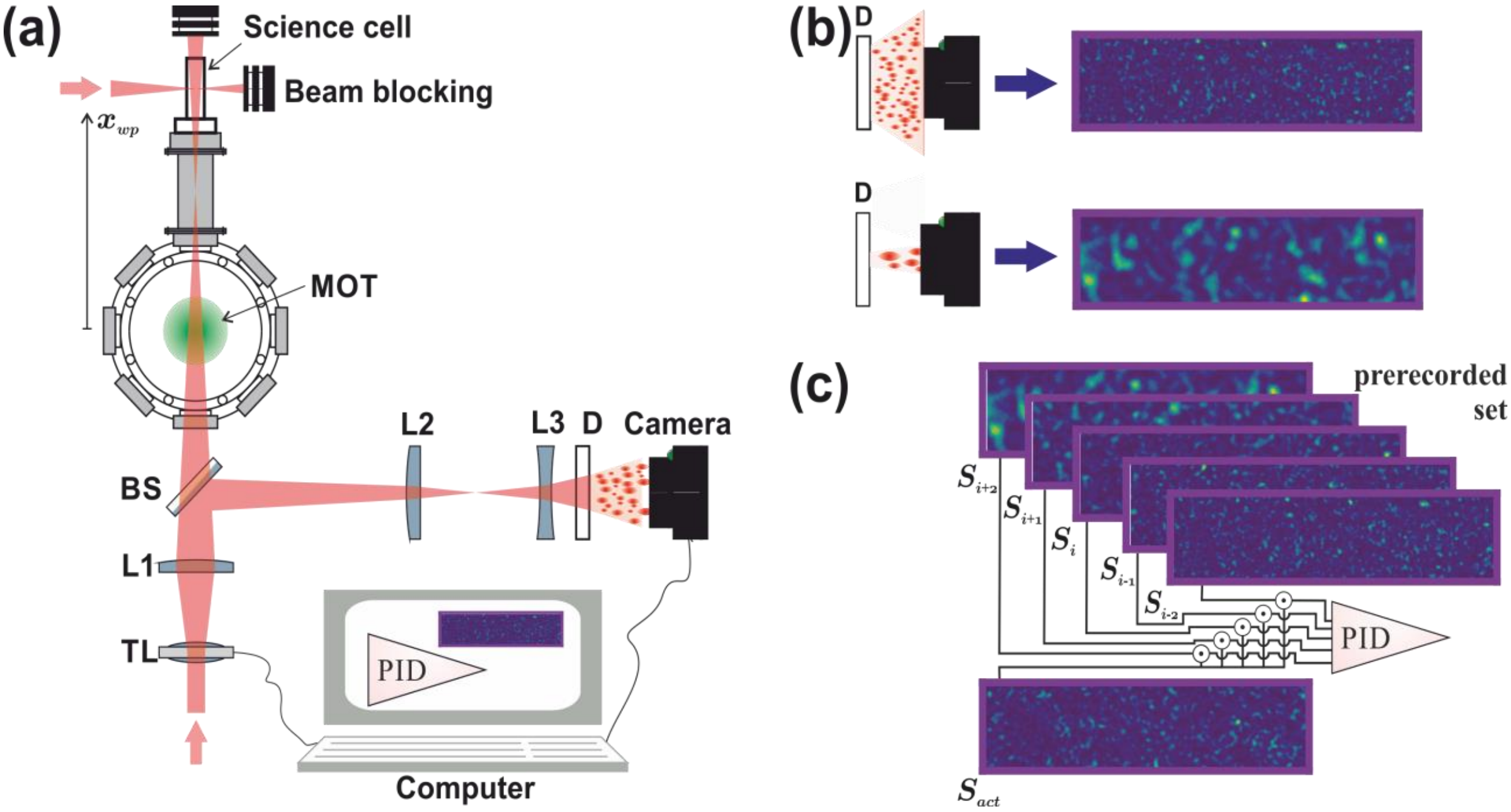

Figure 1 a) Principal optical scheme of transport of atoms from magneto-optical trap to the Science cell. TL – Tunable Lens, L1 – lens with +250 mm focus length, BS – beam splitter, L2 – lens with +500 mm focus length, L3 – lens with −50 mm focus length, D – diffusor or optical rough surface, MOT – magneto optical trap. b) Speckle picture with different size of laser beam on the diffusor. c) Principal scheme of feedback speckle-based algorithm. Here $S_{i\pm n}$, where $n$ from –2 to 2 are the reference pictures, $S_{act}$ is the image captured by camera.

Coherent light beam reflected from a rough surface or passed through a refractive medium with random fluctuations creates a granular interference pattern – speckles. Speckles are used for three-dimensional displacement and temperature measurement [34,35] and in bio- and medical optics [36,37]. The measurement procedure in these applications relies on the fact that a unique speckle pattern corresponds to each value of the measured parameter. By recognizing this pattern, the parameter is unambiguously determined.

If the laser light with a wavelength $\lambda$ a $TEM_{00}$ mode with a radius $w$ scatters on a diffuser, autocorrelation function of the speckle pattern has a gaussian profile and $t$ the width at level $1/e^2$ of this profile $\sigma$ at distance $h$ from the diffuser is given by (see APPENDIX A):

$$\sigma = \frac{\sqrt{2}\lambda h}{\pi w} \tag{1}$$

To measure the speckle pattern of the ODT gaussian beam it was split at the mirror with a polished backside right before the vacuum chamber (BS on Figure 1) and imaged by the CMOS camera. Variation of the dioptric power $D$ of the focus-tunable lens and consequent shift of the beam waist $x_{wp}$ alters the $w$ on the diffuser (Figure 1b) changing the speckle pattern and its width $\sigma$ on the camera.

The focus-tunable lens is driven by a home-made diver based on 16 active bits digital to analog converter with a current $I$, while the $x_{wp}$ of the ODT is determined by the actual dioptric power $D$. Unfortunately, the $D(I)$ dependance can change in time due to heating and other drifts in the lens so the actual $x_{wp}$ is not a perfectly known function of the current. However, the speckle pattern on the camera is determined specifically by the optical properties similarly to the $x_{wp}$ and does not depend on the current to focus position calibration.

Described above speckle properties suggest that dynamic tuning of the driver current $I$ to match the specific speckle pattern $S$ guarantees the specific beam waist position corresponding to this pattern. To implement such a dynamic current tuning the following procedure was developed.

Initially, a reference set of patterns $S_i$ corresponding to a set of currents $I_i$ in the whole spatial range at $x_{wp}$, namely from the magneto-optical trap to the final ODT position, was obtained. The actual waist position produces the speckle pattern $S_{\text{act}}$ on the camera (see Figure 1c). Let's choose a particular $S_{\text{set}}$ from the reference set as a "set value". To determine the offset between the actual position and the set one the dot product $S_{\text{act}} \odot S_i$ is calculated for a range of $S_i$ around $S_{\text{set}}$. As it is seen at the Figure 2a this is a function of the index $i$ that has a maximum attributed to the $S_{\text{act}}$. The value calculated as

$$I_{\text{act}} = \frac{\sum_i I_i \cdot w\,(S_{\text{act}} \odot S_i)}{\sum_i w\,(S_{\text{act}} \odot S_i)} \tag{2}$$

where $w(t) = e^{-|1-t|^3}$, gives the current $I_{\text{act}}$ that can be associated with the actual speckle pattern $S_{\text{act}}$ according to the reference set $\{S_i, I_i\}$.The $w(t)$ function smothers the noise of the curve $S_{\text{act}} \odot S_i$. The value $I_{\text{act}} - I_{\text{set}}$ can be interpreted as an error in a standard PID algorithm to produce the control current $I_{\text{fb}}(t)$ every time the value (2) is measured. The set current can also be a function of time $I_{\text{set}}(t)$.

The actual discretization of the $I_{\text{set}}(t)$ and the corresponding $I_{\text{fb}}(t)$ functions are limited by the inertia of the liquid lens internal fluid. The technical documentation provides the rise time for a given range which estimates the fastest available control rate. If the derivative $\frac{dI_{\text{set}}}{dt}$ sufficiently exceeds this rate, the response to the control current would have a time delay. However, such delay does not necessarily present an issue from the transfer efficiency point of view.

The feedback rate $I_{\text{act}}(t)$ is bounded by the time of obtaining speckle image $S_{\text{act}}$ and the computation time of the value (2), both of which scale with the image size. At the same time, the image size should not be less than the $\sigma$ of the speckle pattern unless the scalar product $S_{\text{act}} \odot S_i$ would be less sensitive to the index $i$. On the other hand, the $\sigma$ should exceed the pixel size for a proper interpretation of the speckle pattern.

Note that according to the formula (1) the average speckle size can be made proportional to the waist position and, therefore, used as a more suitable linear feedback parameter. However, the finite size of the speckle image increases the standard deviation of the actually measured speckle size making it too noisy for a feedback loop. While it is possible to use maximal image size available by the camera to decrease the deviation from linearity, the large image size increases the time of its obtaining from the camera and limits the $I_{\mathrm{act}}(t)$ rate as noted previously. Actually, we used the 264×64 image size with average 0.15 ms acquisition time, while using the full image size 2448×2048 will increase the acquisition time from 1.5 ms for the camera used. Generally, the described procedure allows to overcome the technical limitations of using the speckle size as a feedback parameter.

The feedback control of axial beam waist position was applied during the transport of thulium atoms described above. Once atoms reached the glass cell, the near zero error value was produced during the operation in the wide range of PID parameters and did not require tedious adjusting. However, the actual position of atomic cloud center of mass shifted during the evaporation. To validate the stability of the waist position, the following set of experiments was performed. After the transport the evaporation in the ODT single beam was performed and a successive ramp till the trap depth $U$ was applied. The cloud initially oscillated (see Figure 2c) then settled at a stationary position depending on the trap depth $U$ (see Figure 2b). The steady position confirms the stabilization working. The shift can be caused by a gradient of an external potential not related to the dipole trap. The most common case of such gradient is caused by the beam tilt relative to the horizontal. To verify this hypothesis, the steady positions versus the trap depth were fitted by the formula

$$\Delta z = \frac{-mg\sin(\alpha)}{2U} z_R^2. \tag{3}$$

describing the shift $\Delta z$ of the minimum ODT potential with depth $U$ related to the tilt on the angle $\alpha$ of the ODT beam with $z_R = \frac{\pi w_0^2}{\lambda}$, $w_0$ is the waist radius, $\lambda$ is the light wavelength. The result angle turned out to be 0.5° that is rather reasonable misalignment.

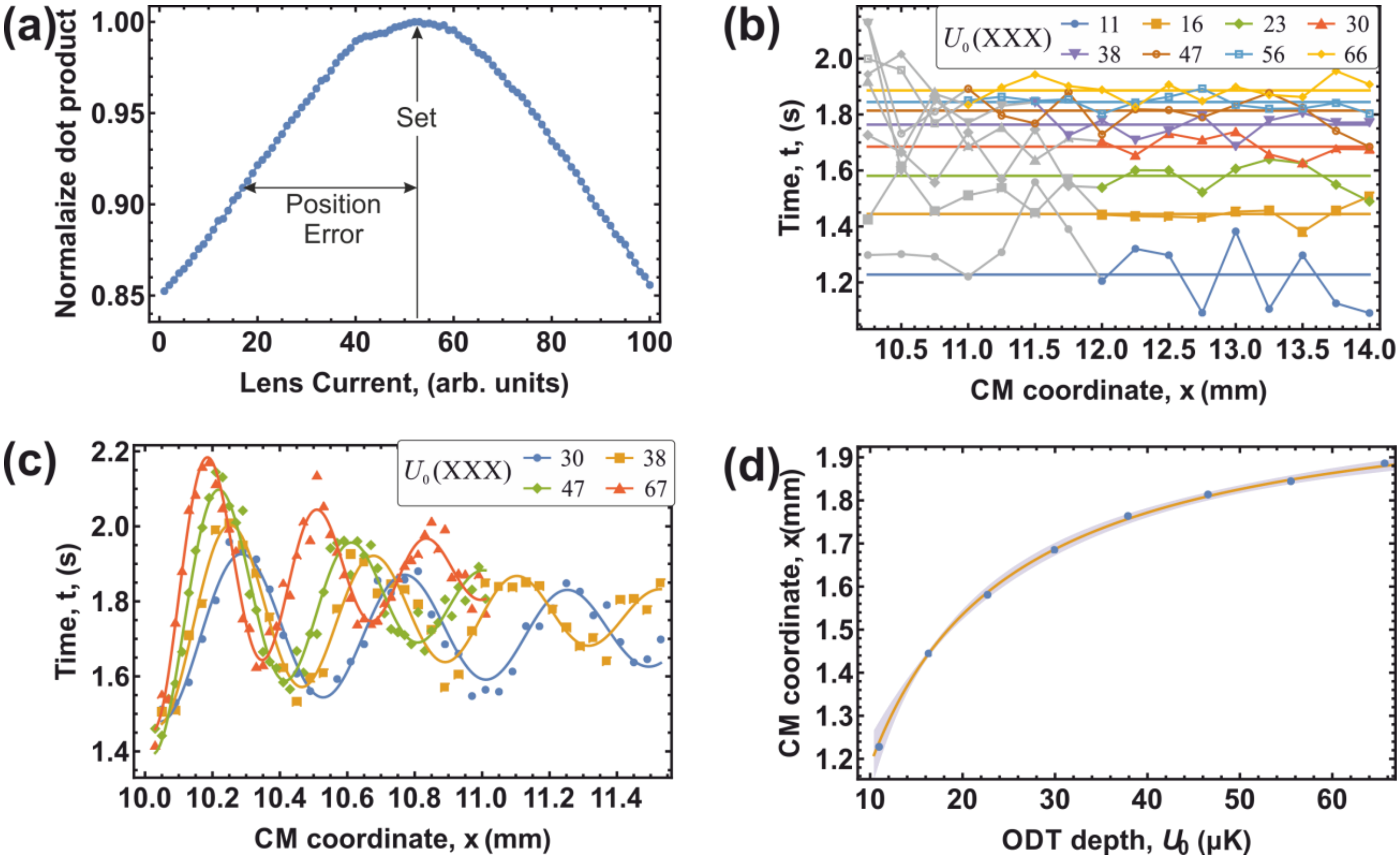

Figure 2 a) The dot product $S_{set} \odot S_i$ for 100 speckle patterns $S_i$ around $S_{set}$. The "Position Error" illustrates the possible deviation of $S_{act}$ from $S_{set}$. b) Dependence of the coordinate of the center of mass (CM) of the atomic cloud along the beam axis on time for different depth of the ODT potential. c) Oscillations of CM for various ODT depth. d) Dependence of the coordinate of the CM of the atomic cloud along the beam axis on the depth of the ODT depth.

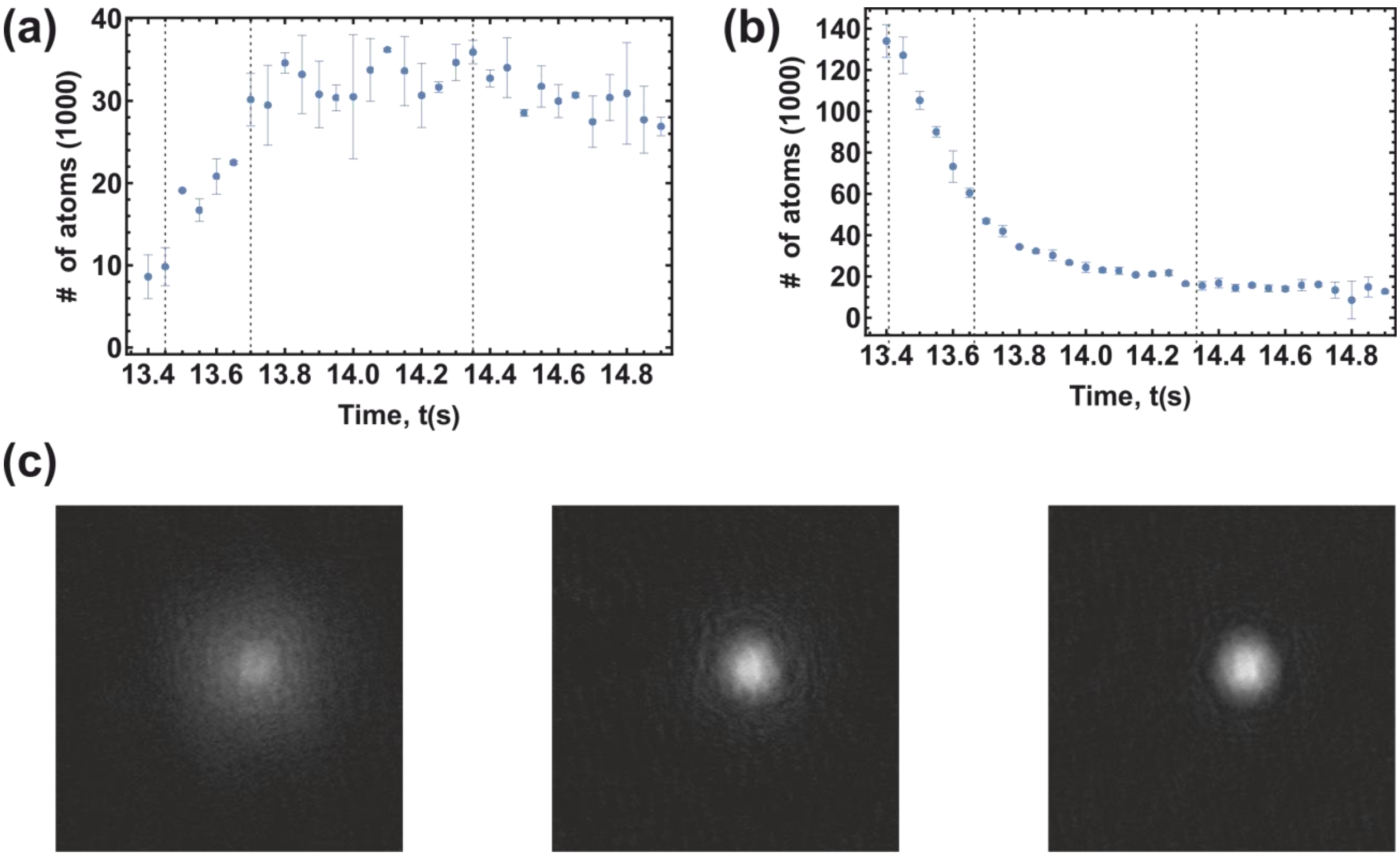

Figure 3. a) The number of condensed atoms versus time of; b) number of atoms in thermal cloud versus time; c) images of atomic cloud at certain points in time, corresponding to the dashed lines and a) and b). Error bars indicate standard deviation.

Exploiting the stabilized transport beam waist, the crossed ODT was created with the second beam, perpendicular to the transport one and the gravity, with a 1064 nm wavelength and 100 μm waist radius. The evaporation cooling in this crossed ODT in the magnetic field 4.8 G produced around $4 \cdot 10^4$ atoms in the Bose-Einstein condensate. The evaporation sequence was recalculated from the one obtained by the optimization procedure [30] regarding the new values of waist for the two beams and simplified to just 3 ramps (compared to 9 in the original one) for a more convenient manual adjustment.

In summary, the stabilization of optical dipole trap longitudinal waist position controlled by a focus tunable lens was realized. The method is based on the feedback-loop comparison of speckle pattern images obtained with the CMOS camera. There is no need of a complete speckle statistics so rather small images can be used, which also accelerates the loop itself. The stabilization provides the transport of cold atoms on 38 cm and the production of the BEC with around $4 \cdot 10^4$ atoms.

## AUTHOR DECLARATIONS

**Daniil Pershin**: Conceptualization (lead), Methodology (lead), Software (lead), Investigation (equal). **Davlet Kumpilov**: Data curation (lead), Investigation (lead), Writing – original draft (lead), Writing – review & editing (equal), Software (supporting), Validation (lead). **Ardjuna Rudnev**: Investigation (equal), Software (equal). **German Subbotin**: Formal analysis (lead), Investigation (equal), Visualization (equal). **Ayrat Ibrahimov**: Formal analysis (equal), Investigation (equal), Software (supporting), Visualization (equal). **Ivan Cojocaru**: Investigation (equal), Software (supporting). **Vladimir Khlebnikov**: Investigation (supporting). **Ivan Pyrkh**: Formal analysis (supporting). **Pavel Aksentsev**: Software (supporting). **Sergey Kuzmin**: Investigation (supporting). **Aleksandr Raskatov**: Investigation (supporting). **Anna Zykova**: Resources (equal). **Vladislav Tsyganok**: Project administration (equal), Supervision (equal). **Alexey Akimov**: Funding acquisition (lead), Project administration (lead), Resources (lead), Supervision (lead), Writing – review & editing (equal).

## ACKNOWLEDGMENTS

This work was supported by Rosatom in the framework of the Roadmap for Quantum computing (Contract No. 868/1759-D dated October 3, 2025).

## APPENDIX A: SPECKLE PATTERN OF A GAUSSIAN BEAM ON A DIFFUSOR

Let us consider the scattering of a Gaussian beam by a diffuser.

The field at an observation point $(x,y)$ is a superposition of spherical waves emitted from diffuser points $(u,v)$ with random phase $\phi(u,v)$:

$$E_{total}(x,y)=\int_{R_D}\frac{A(u,v)}{r(u,v,x,y)}e^{-i(\omega t-kr(u,v,x,y)+\phi(u,v))}K(u,v,x,y)\,dudv\,, \tag{4}$$

Here $A(u,v)$ – is the local complex amplitude and $r(u,v,x,y)\equiv r(u,v)=\sqrt{h^2+R^2}$ – is the distance from the diffuser to the observation point. The phase $\phi(u,v)$ is assumed uniformly distributed on $[-\pi,\pi]$, representing the diffuser's inhomogeneity. The obliquity factor $K(\theta)=\frac{i}{\lambda}\frac{1+cos\theta}{2}$ is neglected in the considered geometry (diffuser–camera distance ~100 mm, camera aperture ~10 mm), where the paraxial approximation is valid.

The intensity $I(x,y)$ is:

$$I(x,y)=\int\limits_{R_D}\int\limits_{R_D}\frac{A^*(u_1,v_1)A(u_2,v_2)}{r(u_1,v_1)r(u_2,v_2)}e^{-ik[r(u_1,v_1)-r(u_2,v_2)]+i[\phi(u_2,v_2)-\phi(u_1,v_1)]}dS_1dS_2. \quad (5)$$

The intensity autocorrelation $\mathcal{C}_{\mathcal{I}}(\gamma,\delta)$ in the screen plane is characteristic speckle size:

$$\mathcal{C}_{\mathcal{I}}(\gamma,\delta)=\frac{I(x,y)I(x+\gamma,y+\delta)}{I^2(x,y)}, \quad (6)$$

Here we assume maximum at $(0,0)$ with decay over the correlation length. For tractability, first consider the field autocorrelation:

$$\mathcal{C}_{\mathcal{E}}(\gamma,\delta)=\frac{E^*(x,y)E(x+\gamma,y+\delta)}{|E|^2(x,y)}. \quad (7)$$

Then,

$$\left\langle E^*(x,y)E(x+\gamma,y+\delta)\right\rangle=\left\langle\int\limits_{R_D}\int\limits_{R_D}\frac{A_1^*A_2}{r_1\tilde{r}_2}e^{-ik[r_1-\tilde{r}_2]+i[\phi_1-\phi_2]}dS_1dS_2\right\rangle \quad (8)$$

where 1 and 2 correspond to the sets of integration variables $(u_1,v_1)$ and $(u_2,v_2)$, $\tilde{r}_k=\sqrt{h^2+(x+\gamma-u_k)^2+(y+\delta-v_k)^2}$.

Because the phase is rapidly oscillating, expand $r$ to zeroth order in denominators and first order in exponents:

$$r=\sqrt{h^2+R^2}\approx h\left(1+1/2\frac{R^2}{h^2}\right)=h\left(1-\frac{xu+yv}{h^2}+\frac{x^2+y^2}{2h^2}+\frac{u^2+v^2}{2h^2}\right).(\text{ SEQ eq1})$$

Finally, we use the delta-correlation model [38]: $\left\langle A_1^*A_2e^{i[\phi_1-\phi_2]}\right\rangle=s\cdot I\delta(u_1-u_2,v_1-v_2)$ with $s$ of order length squared.

So,

$$\left\langle E^*(x,y)E(x+\gamma,y+\delta)\right\rangle=se^{i\varphi}\int\limits_{R_D}\frac{I(u,v)}{h^2}e^{-i\frac{k}{h}[u\gamma+v\delta]}dS, \quad (9)$$

where $\varphi=-\gamma^2-2x\gamma-\delta^2-2y\delta$. For gaussian beam $I(u,v)=I_0e^{-2\frac{u^2+v^2}{w^2}}$, so

$$\left\langle E^*(x,y)E(x+\gamma,y+\delta)\right\rangle = se^{i\varphi}\frac{I_0}{h^2}e^{-w^2k^2\frac{\gamma^2+\delta^2}{8h^2}}\int_{R_D} e^{-2\frac{(u-i\frac{w^2k}{2h}\gamma)^2+(v-i\frac{w^2k}{2h}\delta)^2}{w^2}}dS\,. \quad (10)$$

Following Wick's theorem, $\mathcal{C}_{\mathcal{I}}(\gamma,\delta)=|\mathcal{C}_{\mathcal{E}}(\gamma,\delta)|^2$, therefore:

$$\mathcal{C}_{\mathcal{I}}(\gamma,\delta)=\frac{I_0^2}{P}e^{-w^2k^2\frac{\gamma^2+\delta^2}{4h^2}}\cdot|\mathcal{J}|^2 \quad , \quad (11)$$

$$\mathcal{J}=\int_{R_D} e^{-2\frac{(u-i\frac{w^2k}{2h}\gamma)^2+(v-i\frac{w^2k}{2h}\delta)^2}{w^2}}dS\,, \quad (12)$$

where $P$ is the beam power on the diffusor.

For little size of beam on the diffusor ($R_D \gg \frac{w^2k}{2h}R_C$, $R_C$ – camera aperture radius) the last integral is approximately constant. Hence the correlation function is may be written in the form of gaussian function:

$$\mathcal{C}_{\mathcal{I}}(\gamma,\delta)=\frac{I_0^2}{P}|\mathcal{J}|^2\cdot e^{-\frac{\gamma^2+\delta^2}{2\sigma^2}}\sim e^{-\frac{\gamma^2+\delta^2}{2\sigma^2}}\,. \quad (13)$$

Where we use $k=2\pi/\lambda$ and

$$\sigma=\frac{\sqrt{2}\lambda h}{\pi w} \quad (14)$$

Is its width at $e^2$ level.